\documentclass[conference]{IEEEtran}
\IEEEoverridecommandlockouts

\usepackage{cite}
\usepackage{amsmath,amssymb,amsfonts}
\usepackage{graphicx}
\usepackage{textcomp}
\usepackage{xcolor}
\usepackage{booktabs}
\usepackage{array}
\usepackage{float}
\usepackage{tikz}
\usetikzlibrary{shapes,arrows,positioning,calc}
\usepackage{url}

\def\IEEEtitletopspaceextra{18pt}
\title{AGORA: Can Deliberation and Governance Gates Absorb Participation Bias in Transit Planning?}

\author{
	\parbox{\textwidth}{%
		\centering
		Jung-Hoon Cho$^{1}$, Cathy Wu$^{1,2}$%
	}%
	\thanks{$^{1}$Department of Civil and Environmental Engineering and Laboratory for Information \& Decision Systems, Massachusetts Institute of Technology, Cambridge, MA, USA. (jhooncho@mit.edu, cathywu@mit.edu)}%
	\thanks{$^{2}$Institute for Data, Systems, and Society, Massachusetts Institute of Technology, Cambridge, MA, USA.}%
}

\begin{document}
\maketitle

\begin{abstract}
Transit network design depends not only on the optimization algorithm but also on who shows up to the public hearing. Current practice often collects one-directional comments from self-selected attendees, leaving participant mix as an uncontrolled source of outcome variation. We present AGORA, a framework that holds the network, demand, and solver fixed while systematically varying meeting composition through stakeholder agents, structured deliberation, and governance gates. Across two standard benchmark networks at different scales, we find that (i) aggregate outcomes vary little across compositions, but on tail risk and fairness disparity, representative sampling still tends to outperform skewed compositions; (ii) without deliberation, composition produces no variation at all, showing that deliberation is the mechanism through which who attends affects outcomes; and (iii) governance gates compress cross-profile variance without shifting the average outcome on Mandl, but low acceptance on Mumford0 shows thresholds require instance-specific calibration. These findings reframe participation bias from an uncontrollable input to a process-design problem: even without guaranteed representative attendance, well-structured deliberation and governance criteria can substantially reduce how much outcomes depend on who is in the room.
\end{abstract}


\section{Introduction}
The Transit Network Design Problem ({TNDP}) has a mature optimization literature: given a transit graph, Origin-Destination (OD) demand matrix, and feasibility constraints, select a route set that balances passenger cost, operator cost, and service coverage \cite{farahani2013review,mandl1980evaluation,mumford2013new}. In practice, however, which feasible design is ultimately adopted is mediated by participatory processes (public meetings, comment periods, board votes), where \emph{who shows up} shapes what is judged acceptable. Empirical evidence consistently shows that participation is systematically skewed toward older, higher-income, and car-owning residents \cite{einstein2019participates,einstein2020neighborhood}, a pattern Arnstein's ladder of citizen participation identifies as a structural barrier to equitable outcomes \cite{arnstein1969ladder}. Transport-justice scholarship further argues that fair transit systems require explicit institutional treatment of distributional burdens, not only aggregate efficiency \cite{martens2016transport}. The technical plan that emerges from a public process can therefore depend on meeting composition, even when the underlying travel demand is unchanged.

Current practice illustrates this vulnerability. In the Massachusetts Bay Transportation Authority ({MBTA}) Bus Network Redesign, the agency held virtual public hearings during which attendees were given a few minutes each to comment on a draft route plan \cite{mbta2022bnrd}. Elected officials spoke first; the remaining time was allocated to self-selected residents, most of whom advocated retaining their own bus route. The process provided no structured mechanism for participants to evaluate trade-offs across competing objectives, no explicit equity acceptance criteria, and no way to assess whether a different group of attendees would have produced different policy signals. This format combines a self-selected sample with opaque synthesis, leaving meeting composition as an uncontrolled source of outcome variation.

As AI-assisted decision support enters transportation agencies, the stakeholders who interact with, constrain, or override algorithmic recommendations become a design parameter in their own right. Recent Large Language Model ({LLM})-agent approaches to participatory planning \cite{zhou2024large,yu2024synthetic} and transportation-focused alignment/behavior studies \cite{yan2025addressing,liu2025aligning,wang2025agentic,sun2025llm} demonstrate that profile-conditioned agents can represent policy preferences and adaptive travel behavior at scale, but they do not isolate composition effects under fixed network and demand conditions. Existing frameworks generally do not treat meeting composition as a controlled experimental variable. 

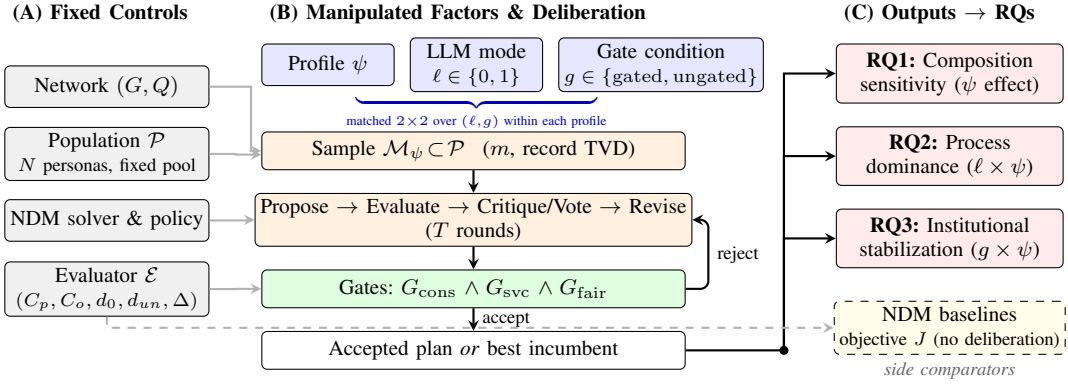
\begin{figure*}[!t]
\centering
\small
\begin{tikzpicture}[
    >=latex,
    node distance=0.35cm and 0.3cm,
    fixedbox/.style={draw, rounded corners=2pt, fill=gray!12,
        minimum width=2.7cm, minimum height=0.55cm,
        font=\footnotesize, align=center, inner sep=2pt},
    manipbox/.style={draw, rounded corners=2pt, fill=blue!10,
        minimum width=1.75cm, minimum height=0.55cm,
        font=\footnotesize, align=center, inner sep=2pt},
    loopbox/.style={draw, rounded corners=2pt, fill=orange!12,
        minimum height=0.50cm,
        font=\footnotesize, align=center, inner sep=2pt},
    gatebox/.style={draw, rounded corners=2pt, fill=green!12,
        minimum height=0.50cm,
        font=\footnotesize, align=center, inner sep=2pt},
    outbox/.style={draw, rounded corners=2pt, fill=red!8,
        minimum width=3.0cm, minimum height=0.55cm,
        font=\footnotesize, align=center, inner sep=3pt},
    refbox/.style={draw, dashed, rounded corners=2pt, fill=yellow!10,
        minimum width=2.6cm, minimum height=0.50cm,
        font=\footnotesize, align=center, inner sep=2pt},
    panellabel/.style={font=\footnotesize\bfseries, anchor=north west},
    arr/.style={->, thick, >=stealth},
    darr/.style={->, thick, >=stealth, dashed},
    jdot/.style={circle, fill=black, inner sep=1.2pt},
]

\node[panellabel] (labA) at (0, 0) {(A) Fixed Controls};
\node[fixedbox, below=0.45cm of labA.south west, anchor=north west] (net)
    {Network $(G, Q)$};
\node[fixedbox, below=0.25cm of net] (pop)
    {Population $\mathcal{P}$\\{\scriptsize $N$ personas, fixed pool}};
\node[fixedbox, below=0.24cm of pop] (solver)
    {NDM solver \& policy};
\node[fixedbox, below=0.26cm of solver] (eval)
    {Evaluator $\mathcal{E}$\\{\scriptsize $(C_p,C_o,d_0,d_{un},\Delta)$}};

\node[panellabel] (labB) at (3.4, 0) {(B) Manipulated Factors \& Deliberation};
\node[manipbox, below=0.15cm of labB.south west, anchor=north west] (prof)
    {Profile $\psi$};
\node[manipbox, right=0.2cm of prof] (llm)
    {{LLM} mode\\{\scriptsize $\ell \in \{0, 1\}$}};
\node[manipbox, right=0.2cm of llm] (gate)
    {Gate condition\\{\scriptsize $g\in\{\mathrm{gated},\mathrm{ungated}\}$}};

\node[font=\scriptsize, below=-0.10cm of llm.south, anchor=north, text=blue!70!black]
    (design) {$\underbrace{\hspace{3.2cm}}_{\text{matched } 2{\times}2 \text{ over }(\ell,g)\text{ within each profile}}$};

\node[loopbox, below=0.6cm of prof.south west, anchor=north west,
      minimum width=5.6cm] (meet)
    {Sample $\mathcal{M}_\psi \!\subset\! \mathcal{P}$
     \enspace ($m$, record TVD)};

\node[loopbox, below=0.30cm of meet, minimum width=5.6cm] (delib)
    {Propose $\to$ Evaluate $\to$ Critique/Vote $\to$ Revise
     \\ ($T$ rounds)};

\node[gatebox, below=0.30cm of delib, minimum width=5.6cm] (gates)
    {Gates: $G_{\mathrm{cons}}$
     $\wedge\; G_{\mathrm{svc}}$
     $\wedge\; G_{\mathrm{fair}}$};

\draw[arr] (meet) -- (delib);
\draw[arr] (delib) -- (gates);

\coordinate (loopR) at ([xshift=0.3cm]gates.east);
\draw[arr, rounded corners=3pt] (gates.east) -- (loopR)
    |- node[font=\scriptsize, near start, right] {reject} (delib.east);

\node[loopbox, below=0.30cm of gates, minimum width=5.6cm, fill=white] (out)
    {Accepted plan \emph{or} best incumbent};
\draw[arr] (gates) -- node[font=\scriptsize, right] {accept} (out);

\draw[arr, gray!60] (net.east) -- ++(0.55,0) |- (meet.west);
\draw[arr, gray!60] (pop.east) -- ++(0.35,0) |- ([yshift=-0.05cm]meet.west);
\draw[arr, gray!60] (solver.east) -- ++(0.55,0) |- (delib.west);
\draw[arr, gray!60] (eval.east) -- ++(0.35,0) |- (gates.west);

\node[panellabel] (labC) at (11, 0) {(C) Outputs $\to$ RQs};

\node[outbox, below=0.15cm of labC.south west, anchor=north west] (rq1)
    {\textbf{RQ1:} Composition\\sensitivity ($\psi$ effect)};
\node[outbox, below=0.30cm of rq1] (rq2)
    {\textbf{RQ2:} Process\\dominance ($\ell \times \psi$)};
\node[outbox, below=0.30cm of rq2] (rq3)
    {\textbf{RQ3:} Institutional\\stabilization ($g \times \psi$)};

\node[refbox, below=0.43cm of rq3] (ref)
    {NDM baselines\\{\scriptsize objective $J$ (no deliberation)}};
\node[font=\scriptsize\itshape, below=-0.02cm of ref.south, text=black!60]
    {side comparators};

\coordinate (junc) at ([xshift=1.3cm]out.east);
\draw[thick] (out.east) -- (junc);
\node[jdot] at (junc) {};
\draw[arr] (junc) |- (rq1.west);
\draw[arr] (junc) |- (rq2.west);
\draw[arr] (junc) |- (rq3.west);

\draw[darr, gray!60] (eval.south) -- ++(0,-0.11) |- (ref.west);
\end{tikzpicture}
\caption{\textbf{AGORA framework}. \textbf{(A)}~Fixed controls held constant across all conditions. \textbf{(B)}~Manipulated factors include meeting composition~$\psi$ and a factorial over {LLM} mode $\ell \in \{0,1\}$ and gate condition within each profile. The deliberation loop iterates propose--evaluate--critique/vote--revise rounds with acceptance gates. \textbf{(C)}~Outputs mapped to research questions; Network Design Method (NDM) baselines serve as side comparators outside the deliberation.}
\label{fig:framework}
\end{figure*}

This paper addresses that gap by posing participation as a counterfactual process question: \emph{how much would planning outcomes change if the same process were run with different meeting compositions?} Real-world planning cannot be rerun under controlled participation regimes. AGORA (\textbf{A}gentic \textbf{G}overnance for \textbf{O}ptimization-\textbf{R}epresentation \textbf{A}lignment, Figure~\ref{fig:framework}) provides this capability: it holds the benchmark instance, demand, and solver fixed while varying only meeting composition, and constructs a normatively richer process than one-directional comment collection, with structured deliberation, explicit trade-off evaluation, and governance gates.

We address three research questions.
\textbf{(RQ1) Composition sensitivity:} under fixed network, demand, and solver, how much do planning outcomes vary when meeting composition changes systematically?
\textbf{(RQ2) Process dominance:} is deliberation or composition the primary driver of outcome variation, and how does the deliberation effect differ across problem scales?
\textbf{(RQ3) Institutional stabilization:} do governance gates reduce residual composition sensitivity, and at what cost to central tendency?
Together, these questions test whether \emph{process design} (deliberation and gates) or \emph{meeting composition} (who is in the room) is the dominant source of outcome variation in AI-assisted participatory planning.

\textbf{Contributions.}
(1)~We quantify the sensitivity of transit network design outcomes to meeting composition, providing evidence that separates participation effects from network topology, demand structure, and solver choice across two benchmark instances of different scales.
(2)~We formalize governance gates with disparity-based acceptance criteria for consensus, coverage, and distributional equity, paired with a matched ungated arm, to quantify how institutional safeguards interact with composition effects.
(3)~We define a paired no-{LLM} pass-through control under identical composition and gate conditions; this arm disables critique-generated revision and post-critique adaptation, so the contrast estimates the bundled effect of the deliberative protocol.

Section~\ref{sec:related} reviews related work. Section~\ref{sec:problem} formulates {TNDP}, Section~\ref{sec:method} presents AGORA, and Section~\ref{sec:experiments} details the experimental design. Section~\ref{sec:results} reports results and Section~\ref{sec:discussion} discusses interpretation.

\section{Related Work}\label{sec:related}

\subsection{Transit Network Design Problem ({TNDP})}
{TNDP} has been extensively studied with route-set decision variables and standardized benchmark instances, most notably the Mandl Swiss network \cite{mandl1980evaluation} and the Mumford family (30--127 nodes) \cite{mumford2013new}, optimizing passenger and operator costs under hard feasibility constraints \cite{guihaire2008transit,farahani2013review}. Methodologically, the literature spans heuristic and evolutionary route-set search \cite{chew2013genetic}, robust/game-theoretic design \cite{laporte2010game,liu2020multiperspective}, data-driven optimization at scale \cite{bertsimas2021data}, and learning-assisted improvement \cite{holliday2025learning}. These formulations enable controlled algorithmic comparisons but generally do not model how participatory processes affect which feasible designs are ultimately selected. Since {TNDP} objectives inherently conflict, the Pareto front contains many feasible designs and the choice among them is a value judgment that optimization alone cannot resolve \cite{arbex2015efficient}. This makes the problem particularly susceptible to participation effects: different stakeholder compositions may favor different trade-off regions. AGORA bridges this gap by asking whether process design can mitigate the influence of meeting composition.

\subsection{Transit equity and fairness}
Transportation planning literature documents systematic accessibility and service disparities across income, disability, and geographic groups in public transit \cite{pereira2017distributive,martens2016transport}. Equity metrics, such as distributional service gaps and accessibility for underserved populations, have been proposed as complements to aggregate efficiency measures \cite{pereira2017distributive}, while planning practice guidance increasingly emphasizes inclusive participation as part of equity accountability \cite{franklin2023inclusive}. However, these metrics are typically applied ex post to finished plans rather than embedded as in-process acceptance criteria. 
AGORA operationalizes disparity bounds as hard acceptance gates, so that the criteria constrain outcomes before a plan is finalized.

\subsection{Participation bias and deliberative planning}
Empirical evidence shows participatory venues systematically over-represent certain demographics and thereby bias policy inputs \cite{einstein2019participates,einstein2020neighborhood,franklin2023inclusive}. Deliberative democracy and collaborative planning traditions emphasize that process design shapes both legitimacy and substantive outcomes \cite{fishkin2009people,innes2010planning}. Arnstein's ladder of citizen participation \cite{arnstein1969ladder} provides a normative framework identifying how nominal inclusion can mask effective exclusion. While prior work documents biased participation, AGORA asks whether process design can mitigate the downstream effects of that bias on planning outcomes.

\subsection{{LLM} agents for deliberation simulation}
{LLM}-based agents can generate role-conditioned critiques and votes, enabling scalable simulation of stakeholder discourse \cite{park2023generative,gao2024large,xi2025rise,liu2025toward}. Persona conditioning is central to this literature \cite{tseng2024two}, but recent evidence highlights substantive validity risks: implicit reasoning bias \cite{gupta2023bias}, weak or unstable demographic instruction effects \cite{magnossao2025effects}, and persona-consistency drift across turns \cite{abdulhai2025consistently}.
Recent orchestration work shows that multi-agent coordination protocols themselves affect task performance and reliability \cite{du2024improving,du2025multi,kim2025towards}. This suggests that deliberation is not merely a channel through which preferences are expressed but an active source of outcome variation.

For transportation, agentic systems are emerging for scenario generation \cite{yu2024synthetic,yu2025preparing}, traffic reasoning \cite{zhang2024trafficgpt,li2024chatsumo,jeong2025agentsumo}, and alignment with travel behavior \cite{yan2025addressing,liu2025aligning,wang2025agentic,sun2025llm,lammer2026gta}. Synthetic-participant construction is also advancing \cite{lim2026large}, and large-scale social simulation platforms \cite{park2024generative,bougie2025citysim} motivate auditable processes when agents are used in policy contexts. In this paper, {LLM}s produce a tractable, structured deliberation protocol with transparent proposal, critique, vote, and revision steps; the goal is controlled sensitivity analysis, not behavioral realism.

\section{Problem Formulation}\label{sec:problem}

\subsection{{TNDP}}
Let $G=(V,E)$ be a transit graph, $Q=[q_{ij}]$ the OD demand matrix, and $K$ the required number of routes.
A design is a route set $X=\{r_1,\dots,r_K\}$, where each $r_k=(v^k_1,\dots,v^k_{|r_k|})$ is a node sequence.

\subsubsection{Hard constraints (C1--C5)}
Let $V(r_k)$ denote the node set of route $r_k$, $H_X$ the transfer graph induced by route set $X$ (route nodes are adjacent if the corresponding routes intersect), and $L_{\min},L_{\max}$ the minimum and maximum permitted route lengths in nodes.
\begin{align}
\textbf{C1 (coverage):} \quad & \bigcup_{k=1}^{K} V(r_k)=V, \\
\textbf{C2 (no cycles):} \quad & v^k_a \neq v^k_b,\ \forall a\neq b,\ \forall k, \\
\textbf{C3 (connected network):} \quad & H_X \text{ is connected}, \\
\textbf{C4 (route length bounds):} \quad & L_{\min}\le |r_k|\le L_{\max},\ \forall k, \\
\textbf{C5 (fixed route count):} \quad & |X|=K.
\end{align}

\subsubsection{{TNDP} metrics}
We evaluate candidate route sets with two cost metrics and two service-quality metrics. The average passenger cost $C_p$ and total operator cost $C_o$ are:
\begin{align}
C_p(X) &= \frac{1}{D} \sum_{i,j} q_{ij}\,c_{ij}(X), \qquad D=\sum_{i,j} q_{ij}, \\
C_o(X) &= \sum_{k=1}^{K} T(r_k),
\end{align}
where $D$ is total demand, $c_{ij}(X)$ is the generalized passenger cost for OD pair $(i,j)$, and $T(r_k)$ is an operator cost proxy (e.g., route travel time).
The \emph{$m$-transfer demand share} $d_m$ and \emph{unserved demand share} $d_{un}$ are:
\begin{align}
    d_m(X)&=\frac{1}{D}\sum_{i,j} q_{ij}\,\mathbf{1}[n_{ij}(X)=m],\quad m\in\{0,1,2\},\\
    d_{un}(X)&=\frac{1}{D}\sum_{i,j} q_{ij}\,\mathbf{1}[\text{unserved or }n_{ij}(X)>2],
\end{align}
where $n_{ij}(X)$ is the minimum number of transfers required. In particular, $d_0$ measures the fraction of demand served without any transfer.
These four metrics have an inherent trade-off: improving coverage ($d_0\!\uparrow$, $d_{un}\!\downarrow$) typically requires longer or more overlapping routes, raising $C_o$, while minimizing $C_p$ may concentrate service on high-demand corridors at the expense of spatial equity. The resulting Pareto front implies that no single route set dominates on all metrics, and selecting among Pareto-efficient designs requires a value judgment that participatory processes are intended to inform.

\subsubsection{Composite objective and baselines}\label{sec:analytical_baselines}
Non-deliberative references use a common scalar objective $J$ with weights $\mathbf{w}=(w_{cp},w_{co},w_{d0},w_{dun})$:
\begin{equation}
    \begin{aligned}
        J(X)=w_{cp}\,C_p(X)+w_{co}\,C_o(X)+w_{d0}\big(1-d_0(X)\big)\\
        \quad +w_{dun}\,d_{un}(X).
    \end{aligned}
    \label{eq:j}
\end{equation}

\textbf{$\ell=0$ process control ($J_\psi$ pass-through).} The $\ell{=}0$ control chooses one seed plan from a fixed catalog by minimizing $J_\psi(X)$ for each meeting composition $\psi$.
\[
J_\psi(X)=\hat w_{cp}\,\hat C_p + \hat w_{co}\,\hat C_o + \hat w_{d0}\,(1{-}\hat d_0) + \hat w_{dun}\,\hat d_{un},
\]
where $\hat{\cdot}$ denotes min-max normalization across the catalog and $(\hat w_{cp},\hat{w}_{co},\hat w_{d0},\hat w_{dun})$ come from the meeting's aggregated preference profile. The selected plan is then evaluated under the same gate logic with no critique-generated revision.

\textbf{Algorithmic Network Design Method (NDM) baselines.} NDM-Equity is a full-demand beam-search baseline with equity-weighted objective terms and is treated as a strong algorithmic reference. NDM-Standard sets $w_{d0}=w_{dun}=0$ in \eqref{eq:j}, prioritizing passenger/operator cost without service-quality or equity terms.

\textbf{Information and objective perturbation variants.} NDM-Partial($k$) restricts NDM-Equity's demand-matrix access to a random fraction $k$ of OD pairs, while NDM-Efficiency and NDM-Equity-Strict shift objective weights toward cost-first and equity-first extremes.

\subsection{Participation bias as a controlled composition shift}
Within a sweep, AGORA holds fixed a synthetic population pool $\mathcal{P}$ and varies the meeting membership $\mathcal{M}$ via profile-conditioned sampling.
Let $\mathcal{P}=\{p_1,\dots,p_N\}$ denote personas with demographic attributes (e.g., income, age, disability, commute mode, home zone, role).
A \emph{meeting profile} $\psi \in \Psi$ induces sampling weights $w_\psi(p)>0$, and a meeting of size $m$ is drawn without replacement, so the sampled meeting $\mathcal{M}_\psi$ is a subset of $\mathcal{P}$ with exactly $m$ participants.

We quantify compositional divergence using Total Variation Distance (TVD). For a categorical attribute, $\mathrm{TVD}(\hat{\pi}_{\mathcal{M}},\hat{\pi}_{\mathcal{P}})=\frac{1}{2}\sum_{g\in\mathcal{G}} \left|\hat{\pi}_{\mathcal{M}}(g)-\hat{\pi}_{\mathcal{P}}(g)\right|$, where $\hat{\pi}_{\mathcal{M}}$ and $\hat{\pi}_{\mathcal{P}}$ are empirical distributions over groups $g$ (e.g., income, commute mode), and $\mathcal{G}$ indexes bins for the attribute being measured.

\subsection{Acceptance gates}\label{sec:gates}
Given a candidate route set $X$, each meeting participant $p_i\in\mathcal{M}$ returns (i) a critique and (ii) a vote $v_i\in[0,1]$. The consensus score is the mean vote, $\bar v(X) = \frac{1}{m}\sum_{p_i\in\mathcal{M}} v_i$.

Fairness disparities are defined over subgroups induced by demographic partitions $\mathcal{A}$ (e.g., income, disability status, commute mode, home zone):
\begin{align}
    \Delta C_p(X) &= \max_{a\in\mathcal{A}} \left| \mu_{C_p}(X \mid a) - \mu_{C_p}(X) \right|,\\
    \Delta d_{un}(X) &= \max_{a\in\mathcal{A}} \left| \mu_{d_{un}}(X \mid a) - \mu_{d_{un}}(X) \right|,\\
    \Delta d_0(X) &= \max_{a\in\mathcal{A}} \left| \mu_{d_0}(X \mid a) - \mu_{d_0}(X) \right|,
\end{align}
where $\mu_{\cdot}(X \mid a)$ denotes a subgroup-conditional mean computed by the evaluation pipeline.

We evaluate three gates with policy thresholds
$\tau_{\mathrm{cons}}, \tau_{\mathrm{svc}}, \tau_{\mathrm{cp}}, \tau_{\mathrm{dun}}, \tau_{\mathrm{d0}}$:
\begin{align}
    G_\mathrm{cons}:& \quad \bar v(X)\ge \tau_{\mathrm{cons}}, \label{eq:g_cons}\\
    G_\mathrm{svc}:& \quad d_{un}(X)\le \tau_{\mathrm{svc}}, \label{eq:g_svc}\\
    G_\mathrm{fair}:& \quad \Delta C_p(X)\le \tau_{\mathrm{cp}} \wedge \Delta d_{un}(X)\le \tau_{\mathrm{dun}} \nonumber\\
    &\quad\quad\quad\quad\quad\quad \wedge \Delta d_0(X)\le \tau_{\mathrm{d0}}. \label{eq:g_fair}
\end{align}
A plan is \emph{accepted} if all gates pass in the same deliberation round; otherwise the system returns the best incumbent under a composite gate-violation penalty.

\section{Method}\label{sec:method}
\subsection{Framework overview}
Each AGORA run proceeds in four stages. First, a synthetic population $\mathcal{P}$ is generated once per sweep and held fixed. Second, a meeting $\mathcal{M}_\psi$ is sampled from $\mathcal{P}$ using profile $\psi$ and a fixed seed. Third, up to $T$ propose--evaluate--critique/vote--revise rounds produce a sequence of candidate route sets $X^{(t)}$. Fourth, acceptance gates \eqref{eq:g_cons}--\eqref{eq:g_fair} are checked each round; the run terminates on the first round where all gates pass, or returns the best incumbent if no round passes.

\subsection{Persona construction}
Each persona $p_i \in \mathcal{P}$ is instantiated from a parameterized archetype template and assigned demographic attributes, preference weights over criteria, and a profile-dependent participation rate. Critiques and votes are elicited via persona-conditioned prompts to a backbone {LLM}, which receives the persona profile, the current route-set description, and the evaluated {TNDP} metrics. This design separates persona specification (fixed attributes and weights) from persona behavior (critique and voting). 

\subsection{NDM solver and deliberation seeding}
NDM is a two-phase constrained heuristic: (i)~demand-weighted greedy construction that iteratively selects routes maximizing a coverage score, followed by (ii)~iterative local search with add, swap, and remove operators under hard feasibility constraints \cite{mumford2013new,farahani2013review}. NDM-Equity uses a separate beam-search procedure with full demand-matrix simulation at each step. NDM-Equity and NDM-Standard are evaluated once per sweep as non-deliberative references. For deliberation, each run is seeded with a $J_\psi$-selector solution: a weighted-sum scalarization over min-max-normalized objectives ($C_p$, $C_o$, $1{-}d_0$, $d_{un}$), where weights are derived from the aggregated persona preferences of the meeting composition $\psi$. The selector chooses from a fixed catalog of published Pareto-front solutions for each benchmark instance \cite{nikolic2013transit,arbex2015efficient}. Under $\ell=0$, no route modification occurs and the $J_\psi$-selector solution passes through as the final outcome.

\subsection{Deliberation protocol}
For each deliberation round $t=1,\dots,T$:
\begin{enumerate}
    \item \textbf{Propose:} a planning agent generates or revises a candidate route set $X^{(t)}$. To keep changes auditable, we enforce a per-round edit budget.
    \item \textbf{Evaluate:} compute {TNDP} metrics and disparity metrics for $X^{(t)}$ against the fixed demand matrix $Q$.
    \item \textbf{Critique \& vote:} each meeting participant $p_i\in\mathcal{M}_\psi$ returns critique text and vote $v_i^{(t)}\in[0,1]$; compute $\bar v^{(t)}$.
    \item \textbf{Revise:} the planning agent updates the proposal based on critiques.
\end{enumerate}
Within each round, AGORA executes up to $S_{\mathrm{rev}}$ inner revision sub-rounds to respond to aggregate critique before finalizing $X^{(t)}$.
This is a \emph{reactive} propose--critique--revise protocol: the agent revises using observed critiques.

\subsection{Paired {LLM}-mode control}
To characterize process effects under matched $(\psi,s,g)$ cells, we run paired modes with $\ell=1$, where {LLM}s generate critiques, votes, and planning revisions, and $\ell=0$, where the $J_\psi$-selected seed is evaluated under the same gate logic but no critique text is generated and no route revision is executed. Within-cell pairing ($\ell=1$ vs $\ell=0$ at fixed $\psi, s, g$) identifies the added effect of enabling critique-mediated revision relative to a no-revision pass-through arm. This is a process-bundle contrast (language + revision + adaptive search), not a language-only causal estimate.

\section{Experiments}\label{sec:experiments}

\subsection{Benchmark instances and meeting profiles}
This study evaluates AGORA on two benchmark instances of different scale: Mandl/Mandl1 \cite{mandl1980evaluation} (15 nodes, 42 directed links, 12 routes) and Mumford/Mumford0 \cite{mumford2013new} (30 nodes, 90 links, 12 routes), both with fixed OD demand $Q$ \cite{arbex2020transitnetworkdesign}. Mandl provides a compact proof-of-concept; Mumford0 tests whether findings generalize to a wider Pareto front and greater spatial heterogeneity. Within each sweep, we fix the population pool at $N=64$ personas, meeting size at $m=8$, deliberation depth at $T=8$ rounds with $S_{\mathrm{rev}}=3$ inner revision sub-rounds, and all evaluator settings constant across conditions.

We evaluate eight meeting profiles $\psi\in\Psi$ that over- or under-weight demographic groups in meeting attendance.
Table~\ref{tab:profiles} summarizes the intended direction of bias and the achieved composition shift, measured as income TVD between the sampled meeting and the full population.

\begin{table}[t]
\centering
\caption{Meeting profiles with composition shift (Income TVD from full population; higher = more skewed).}
\label{tab:profiles}
\vspace{-5pt}
\resizebox{0.96\columnwidth}{!}{%
\begin{tabular}{p{1.7cm}p{2.8cm}p{1.9cm}c}
\toprule
Profile $\psi$ & Over-weighted & Under-weighted & Income TVD \\
\midrule
Representative & stratified quota mix & none & 0.156 \\
\addlinespace[1pt]
Affluent Senior & high income, 55+, car owners & low income & 0.405 \\
\addlinespace[1pt]
Car owner & car commuters, business roles & bus/walk commuters & 0.373 \\
\addlinespace[1pt]
Equity Voices & low income, transit-dep., disability & high income & 0.381 \\
\addlinespace[1pt]
Transit Dependent Residents & low income, bus/walk, low participation & high-income car owners & 0.452 \\
\addlinespace[1pt]
Retired Homeowners & retired, older, mod./high income & low-income workers & 0.225 \\
\addlinespace[1pt]
Downtown Professional & selected zones, professionals & low income & 0.295 \\
\addlinespace[1pt]
Civic Regulars & advocacy/MPO insiders & low engagement & 0.175 \\
\bottomrule
\end{tabular}}
\end{table}

\subsection{Experimental Design}
The experimental unit is one run indexed by $(\psi,s,\ell,g)$, where $\psi \in \Psi$ is the meeting profile, $s$ is one of 10 fixed random seeds shared across all arms, $\ell \in \{0,1\}$ is {LLM} mode ({GPT}-4.1-mini \cite{openai2025gpt41}), and $g \in \{\text{gated}, \text{ungated}\}$ is gate condition.
Network instance, OD demand, population pool, evaluator settings, and route-edit budget are held fixed across all runs.

Matched indexing defines three research contrasts.
\textbf{RQ1 (composition sensitivity):} compare across profiles $\psi$ under fixed $(s,\ell,g)$, pooling over seeds, to quantify cross-profile spread in each metric family.
\textbf{RQ2 (process dominance):} compare composition spread under $\ell=1$ vs $\ell=0$ at fixed $(s,g)$ as a process-bundle contrast. The $\ell=0 \to \ell=1$ difference quantifies variation introduced by enabling critique-mediated revision beyond pass-through seeding.
\textbf{RQ3 (institutional stabilization):} compare composition spread under gated vs ungated at fixed $(s,\ell)$; $\sigma^2_{\text{gated}}/\sigma^2_{\text{ungated}} < 1$ indicates that gates compress residual composition sensitivity.

\subsection{Operational parameters}
We instantiate the gate thresholds in \eqref{eq:g_cons}--\eqref{eq:g_fair} as $\tau_{\mathrm{cons}}=0.55$, $\tau_{\mathrm{svc}}=0.20$, $\tau_{\mathrm{cp}}=10.0$, $\tau_{\mathrm{dun}}=0.25$, and $\tau_{\mathrm{d0}}=0.30$. The composite objective weights in \eqref{eq:j} are $(w_{cp},w_{co},w_{d0},w_{dun})=(2,1,1,4)$, fixed across all reported arms. The $J_\psi$-selector catalog contains 11 published Pareto-front solutions for Mandl \cite{nikolic2013transit,arbex2015efficient}.
\subsection{Statistical analysis}
Each run is indexed by $(\psi,s,\ell,g)$ with matched seeds across profiles and arms. We summarize each arm with median, interquartile range (IQR), and acceptance proportion.

For RQ1 (composition sensitivity), we test profile effects within each benchmark-condition using Kruskal-Wallis and report cross-profile dispersion using the range and coefficient of variation (CV) of profile medians. We estimate the TVD-disparity association as a single pooled Spearman rank correlation across all individual observations within each gate arm.

For RQ2 (process dominance), the estimand is a seed-paired change in cross-profile spread rather than a median shift. For metric $m$, cross-profile spread is $S_{s,g,\ell}^{(m)}=\max_{\psi} m(\psi,s,\ell,g)-\min_{\psi} m(\psi,s,\ell,g)$, and the paired process contrast is $\Delta S_{s,g}^{(m)}=S_{s,g,\ell=1}^{(m)}-S_{s,g,\ell=0}^{(m)}$. We test directionality with exact paired sign tests across seeds.

For RQ3 (institutional stabilization), we use the same spread framework at fixed $(s,\ell)$, with $\Delta S_{s,\ell}^{(m)}=S_{s,\text{gated},\ell}^{(m)}-S_{s,\text{ungated},\ell}^{(m)}$; negative values indicate variance compression under gates. With 10 seeds per profile, $p$-values are interpreted as screening diagnostics and reported alongside effect sizes.

All non-deliberative baselines are evaluated once per sweep under the same {TNDP} evaluator and demand matrix as deliberative runs. {TNDP} metrics are reported for all conditions.

\section{Results}\label{sec:results}
This section summarizes the experimental results on both Mandl and Mumford0 under fixed network, demand, and evaluator settings. Detailed values and statistical outputs are reported in Table~\ref{tab:main_results} and Figures~\ref{fig:composition_mandl}--\ref{fig:guardrail_effect}. We organize the results by the three RQs.

Table~\ref{tab:main_results} reports median outcomes over 10 matched seeds for both benchmarks, with profile-specific rows for the $\ell{=}1$ arms and pooled references for the fixed-output controls ($\ell{=}0$ pass-through and NDM baselines), which produce the same result regardless of meeting composition. Across profiles, medians are tightly clustered, in contrast to the larger differences induced by deliberation and gate condition. Mandl is mostly permissive under the current thresholds, whereas Mumford0 is substantially harder, with much lower gated acceptance; the $\ell{=}0$ control narrowly fails the service gate on Mumford0.
\begin{table}[t]
\centering
\caption{Cross-benchmark results (median over 10 seeds). The $\ell{=}0$ arm is the $J_\psi$-selector pass-through control. The $\ell{=}0$ arm and NDM baselines have identical output across profiles.}
\label{tab:main_results}
\scriptsize\setlength{\tabcolsep}{2.0pt}%
\vspace{-5pt}
\resizebox{0.96\columnwidth}{!}{%
\begin{tabular}{@{}l ccccc ccccc@{}}
\toprule
& \multicolumn{5}{c}{\textbf{Mandl} (15 nodes, 12 routes)}
& \multicolumn{5}{c}{\textbf{Mumford0} (30 nodes, 12 routes)} \\
\cmidrule(lr){2-6} \cmidrule(lr){7-11}
Profile
  & Acc & $C_p\!\downarrow$ & $C_o\!\downarrow$ & $d_0\!\uparrow$ & $d_{un}\!\downarrow$
  & Acc & $C_p\!\downarrow$ & $C_o\!\downarrow$ & $d_0\!\uparrow$ & $d_{un}\!\downarrow$ \\
\midrule
\multicolumn{11}{@{}l}{\textit{$\ell{=}1$, gated ({LLM} deliberation + governance gates)}} \\
\;\;Rep.        & 90\%  & 16.97 & 142 & .789 & .000  & 20\%  & 29.39 & 189 & .223 & .240 \\
\;\;Affluent    & 90\%  & 16.71 & 136 & .799 & .000  & 20\%  & 29.62 & 194 & .224 & .235 \\
\;\;Car-Owner   & 90\%  & 17.01 & 135 & .780 & .000  & 40\%  & 29.18 & 200 & .239 & .202 \\
\;\;Equity      & 90\%  & 16.82 & 139 & .792 & .000  & 20\%  & 29.53 & 201 & .235 & .242 \\
\;\;Transit-Dep.& 90\%  & 16.76 & 134 & .796 & .000  & 30\%  & 29.97 & 195 & .216 & .238 \\
\;\;Retired     & 90\%  & 16.83 & 135 & .797 & .000  & 30\%  & 30.67 & 189 & .210 & .320 \\
\;\;Downtown    & 90\%  & 16.97 & 137 & .781 & .000  & 40\%  & 29.40 & 196 & .229 & .241 \\
\;\;Civic       & 100\% & 17.20 & 129 & .765 & .001  & 50\%  & 29.35 & 191 & .234 & .204 \\[1.5pt]
\midrule
\multicolumn{11}{@{}l}{\textit{$\ell{=}1$, ungated ({LLM} deliberation, no gates)}} \\
\;\;Rep.        & -- & 17.07 & 135 & .769 & .000  & -- & 29.77 & 194 & .224 & .252 \\
\;\;Affluent    & -- & 17.83 & 122 & .720 & .005  & -- & 29.96 & 197 & .230 & .265 \\
\;\;Car-Owner   & -- & 17.09 & 131 & .776 & .001  & -- & 30.49 & 191 & .213 & .288 \\
\;\;Equity      & -- & 17.38 & 132 & .764 & .000  & -- & 29.91 & 189 & .220 & .236 \\
\;\;Transit-Dep.& -- & 17.90 & 131 & .718 & .022  & -- & 30.75 & 189 & .214 & .304 \\
\;\;Retired     & -- & 17.30 & 132 & .757 & .000  & -- & 30.20 & 200 & .226 & .266 \\
\;\;Downtown    & -- & 17.05 & 136 & .781 & .002  & -- & 29.50 & 194 & .229 & .224 \\
\;\;Civic       & -- & 17.17 & 134 & .760 & .000  & -- & 30.10 & 198 & .226 & .272 \\[1.5pt]
\midrule
$\ell{=}0$, $J_\psi$ & 100\% & 15.09 & 312 & .989 & .000  & 0\% & 29.80 & 196 & .221 & .227 \\
\midrule
\multicolumn{11}{@{}l}{\textit{Baselines (deterministic; no deliberation)}} \\
NDM-Equity      & -- & 16.31 & 116 & .838 & .000  & -- & 34.50 & 139 & .155 & .498 \\
NDM-Standard    & -- & 21.55 & 66  & .563 & .167  & -- & 38.78 & 85  & .079 & .766 \\
NDM-Partial(50\%) & -- & 36.13 & 77 & .251 & .714  & \multicolumn{5}{c}{--} \\
\bottomrule
\end{tabular}
}
\end{table}
\subsection{Composition sensitivity (RQ1)}
Aggregate composition effects are not statistically significant at $\alpha{=}0.05$. On Mandl, no profile effect is detected for any metric (Kruskal--Wallis $p>0.52$), and cross-profile CV remains below 3\% for $C_p$, $C_o$, and $d_0$. On Mumford0, this insensitivity holds: all tests remain non-significant and scaling up the network does not amplify composition effects.

\begin{figure}[t]
\centering
\includegraphics[width=0.99\columnwidth]{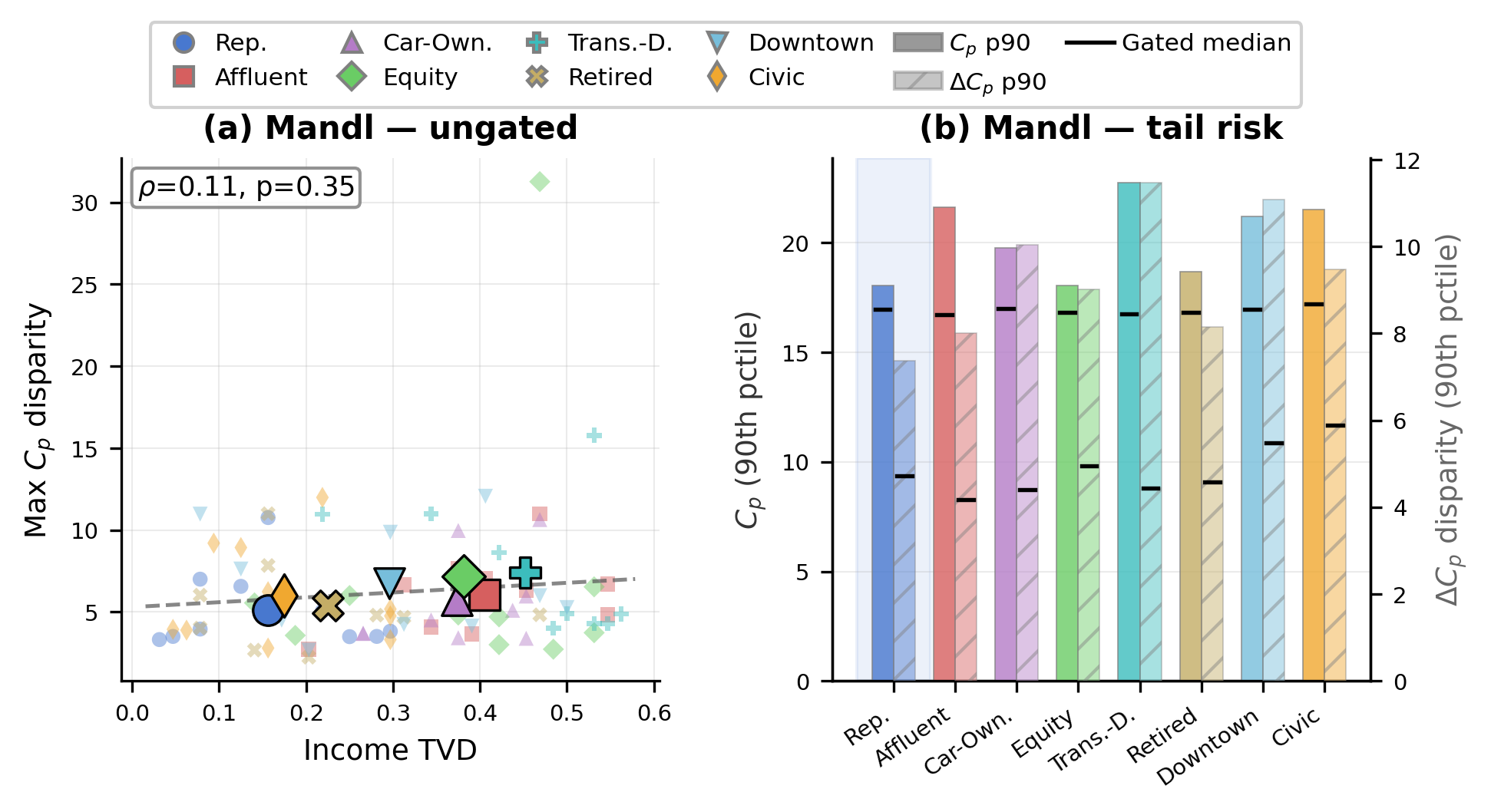}%
\caption{Composition sensitivity (RQ1). (a)~Income TVD vs.\ maximum $C_p$ disparity (ungated, $\ell{=}1$); large markers are profile centroids. (b)~Tail-risk comparison: solid bars show 90th-percentile $C_p$ (left axis), hatched bars show 90th-percentile $\Delta C_p$ disparity (right axis); black ticks mark gated medians. Representative profile (blue highlight) achieves the lowest tail risk on Mandl.}
\label{fig:composition_mandl}
\end{figure}

However, point-estimate differences reveal a consistent directional pattern (Figure~\ref{fig:composition_mandl}a). The Representative profile achieves the lowest maximum passenger-cost disparity, whereas the most skewed profile, Transit-Dep., produces the highest. The Spearman association between income TVD and disparity is positive but non-significant ($\rho{=}0.11$, $p{=}0.35$).

Tail-risk analysis sharpens this pattern (Figure~\ref{fig:composition_mandl}b). Under ungated deliberation on Mandl, the Representative profile achieves the lowest 90th-percentile passenger cost and passenger-cost disparity, while the most skewed profiles produce the worst tail outcomes. A composite scalarized objective $J$ under fixed common weights ranks Representative first and Transit-Dep.\ last in the ungated arm. Governance gates compress these differences substantially: gated tail-risk values are nearly identical across profiles, suggesting that governance gates can reduce this directional pattern and make it manageable through institutional design.

\subsection{Process dominance (RQ2)}
In these runs, the no-revision $J_\psi$ control returns the same catalog solution across profiles. Deliberation ($\ell=1$) is therefore the mechanism through which meeting composition can influence outcomes, and as RQ1 shows, that influence remains small.

\begin{figure}[t]
    \centering
    \includegraphics[width=0.87\columnwidth]{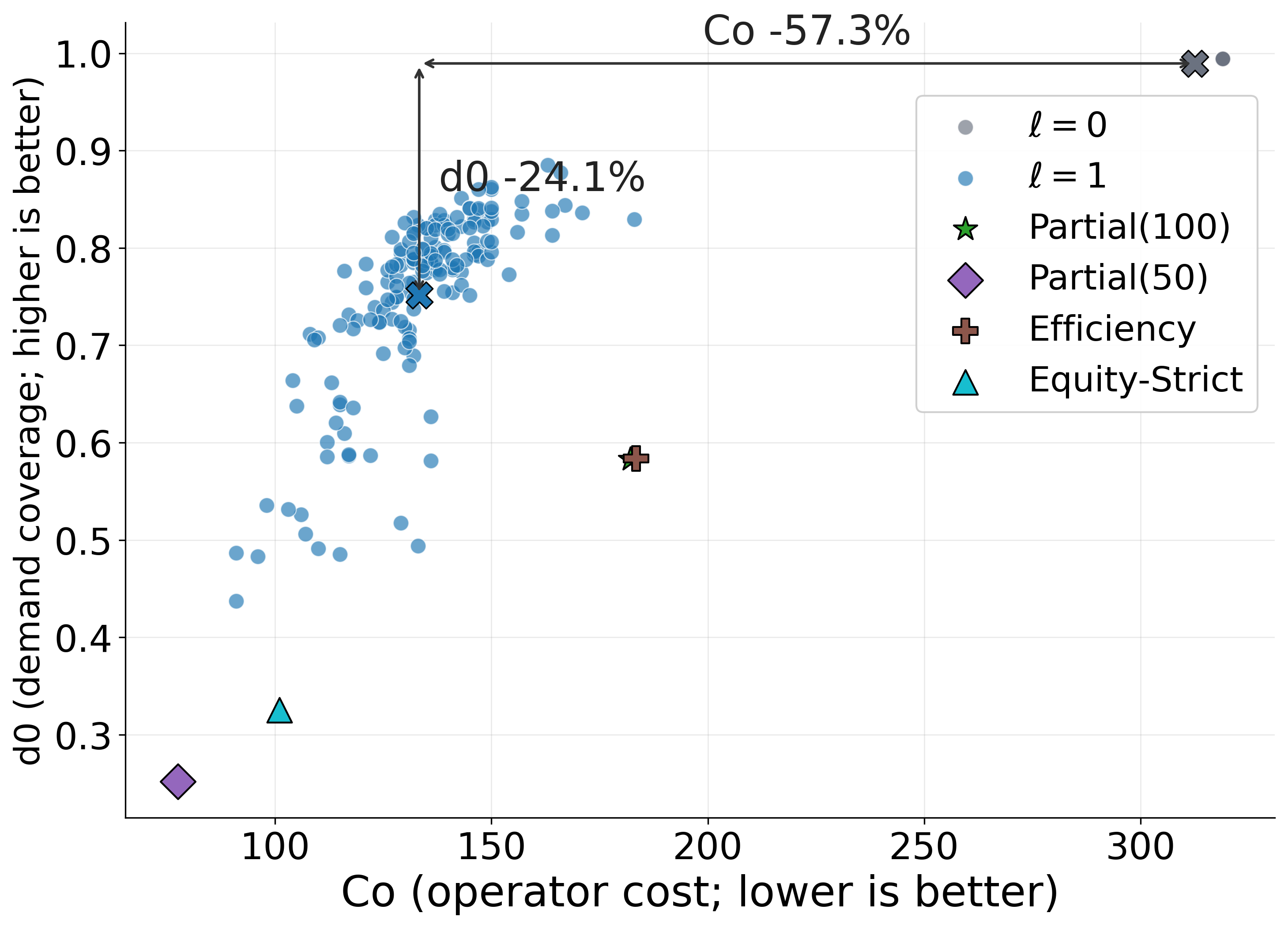}%
    \caption{Process dominance (RQ2): $\ell{=}1$ vs $\ell{=}0$ in the $(C_o, d_0)$ plane on Mandl. The $\ell{=}0$ selector achieves near-perfect coverage at high operator cost; deliberation halves $C_o$ at the expense of $d_0$. NDM information-perturbation baselines shown for reference.}
    \label{fig:paired_effect}
\end{figure}

On Mandl, deliberation creates a rider--operator trade-off: the $\ell=0 \to \ell=1$ shift raises passenger cost while halving operator cost at a smaller cost in direct service (Figure~\ref{fig:paired_effect}). Deliberation rebalances toward operational efficiency. Removing gates amplifies this rebalancing slightly. On Mumford0, the $\ell{=}0$ control already occupies the same region of the objective space as deliberated solutions, so the aggregate quality gap is small; the primary effect of enabling deliberation is to introduce cross-profile variation rather than to shift central tendency. The $\ell{=}0$ control narrowly fails the service gate, while deliberation enables some runs to cross that threshold.

\subsection{Institutional stabilization (RQ3)}
\begin{figure}[t]
\centering
\includegraphics[width=0.97\columnwidth]{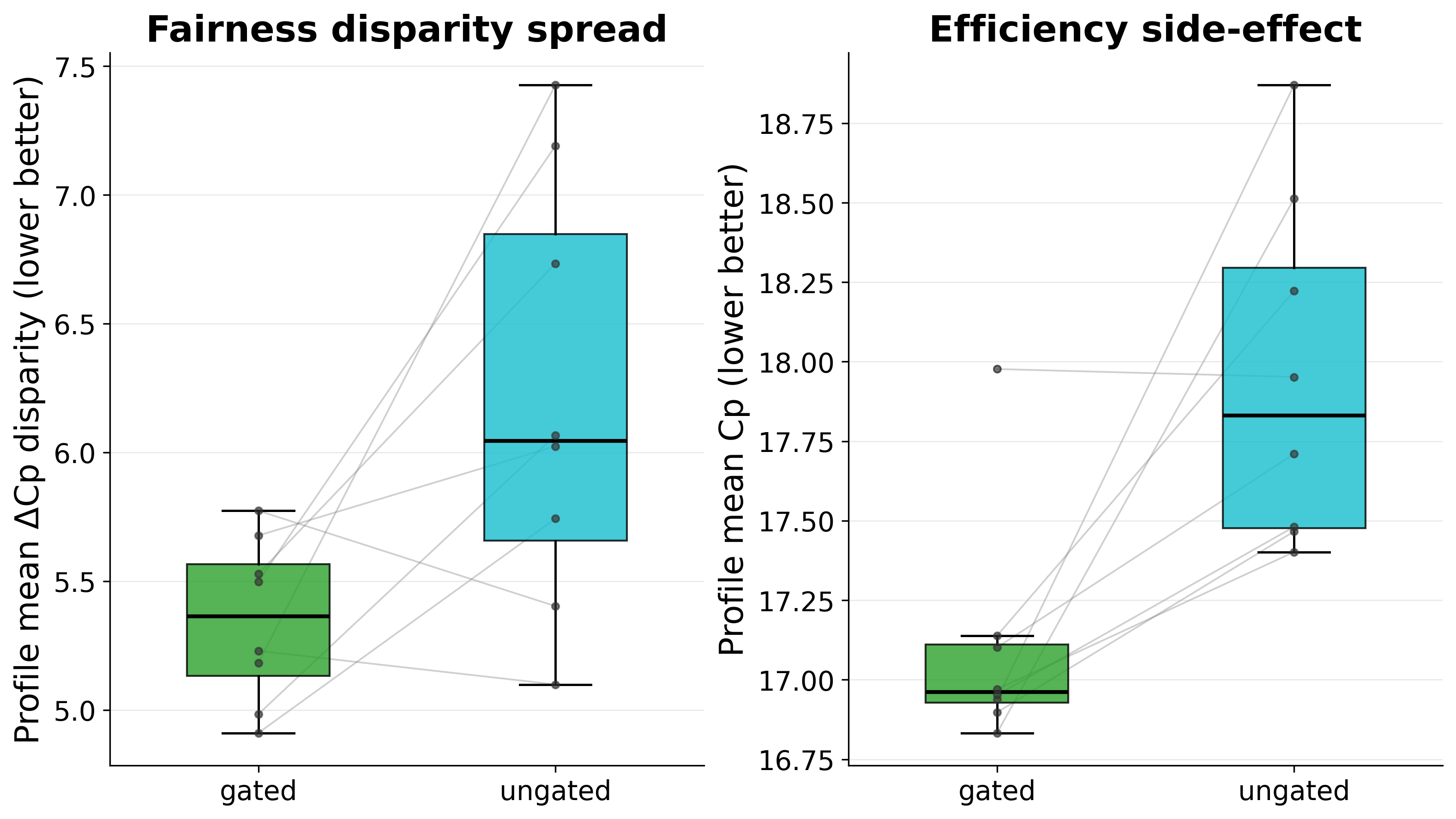}%
\caption{Institutional stabilization (RQ3): governance gates attenuate between-profile disparity variance on the Mandl instance.}
\label{fig:guardrail_effect}
\end{figure}

Governance gates act primarily on variance. On Mandl, gated and ungated medians are nearly identical, but the gate substantially compresses cross-profile spread and reduces within-profile variance by an order of magnitude for most profiles (Figure~\ref{fig:guardrail_effect}). The effect is sharpest for the most composition-sensitive profiles. Gates thus narrow the extent to which participation composition can influence outcomes without altering what the average composition produces.

On Mumford0, gates no longer compress variance; instead, the service gate rejects the majority of runs. This highlights a practical limitation: fixed gate thresholds do not transfer across problem scales. Instance-specific calibration is necessary to avoid either vacuous gates or excessively restrictive ones.

\subsection{Algorithmic baseline comparison}
NDM-Equity, which has full demand-matrix access, remains a strong reference baseline (Table~\ref{tab:main_results}). On Mandl, deliberation approaches NDM-Equity on operator cost while the $\ell=0$ control achieves lower rider cost at a ${\sim}3{\times}$ operator-cost penalty. On Mumford0, the $\ell{=}0$ control and deliberation produce comparable aggregate metrics, both substantially outperforming NDM-Equity on rider cost while incurring higher operator cost. NDM-Partial, which sees only half the demand matrix, performs far worse than both NDM-Equity and deliberation on Mandl, illustrating the sensitivity of heuristic search to the completeness of demand information.

\section{Discussion}\label{sec:discussion}
\textbf{Motivation versus observed sensitivity.} The motivating concern is that participatory forums are systematically skewed and can shape which plans are politically adoptable \cite{einstein2019participates,einstein2020neighborhood}. The RQ1 finding here is narrower: under fixed demand, geometry, and evaluation rules, aggregate outcomes vary little across composition profiles at the benchmark scale. Composition can materially affect who is heard, which concerns are prioritized, and which plans are judged acceptable, even when accepted plans have similar aggregate metrics.

\textbf{Quality--process trade-off.} On Mandl, deliberation rebalances plans toward operator efficiency at the cost of rider metrics, while NDM-Equity dominates on aggregate quality. However, transportation agencies adopt plans through institutional processes that require public hearings, surveys, reasoning, contestability, and accountable documentation \cite{innes2010planning,franklin2023inclusive}. AGORA addresses this by making gate criteria, critique traces, and acceptance rationales explicit and auditable; the relevant comparison is thus not raw solution quality but quality conditioned on procedural accountability.

\textbf{Information structure.} NDM-Equity's strong performance reflects its structural information advantage rather than access to the correct objective weights. We interpret it as a strong reference, not a proven optimum. On Mandl, the $\ell{=}0$ catalog selector achieves near-optimal rider cost at high operator cost; deliberation rebalances this trade. On Mumford0, the catalog selector already reaches a comparable region of the objective space, so deliberation's aggregate contribution is marginal.

\textbf{Scale and cross-benchmark comparison.} Mandl's compact feasible region limits observable spread; Mumford0 provides a first scale-up test. Three cross-benchmark contrasts are instructive: (i)~composition sensitivity does not amplify with network size; (ii)~gate-binding rates diverge sharply, indicating that fixed thresholds do not transfer across scales; and (iii)~variance compression weakens when profile-level variance is already tight. On both benchmarks the $\ell{=}0$ control produces competitive aggregate metrics, but the deliberation-vs-control gap is larger on Mandl (where deliberation halves operator cost) than on Mumford0 (where the gap is marginal). For real-city networks with greater spatial and demand heterogeneity, composition effects may be stronger.

\textbf{Representative sampling as a preferred default.} While no single profile produces statistically significantly better aggregate outcomes, directional evidence in these benchmarks consistently favors representative sampling on disparity and tail-risk metrics. The Representative profile achieves the lowest cross-zone disparity, ranks first on the composite objective, and performs best in the reported Mandl tail-risk analysis; the broader comparisons also directionally favor representative sampling. The most composition-skewed profiles consistently rank worst. When agencies cannot guarantee balanced attendance, deploying representative synthetic populations provides a conservative safeguard---and governance gates compress any residual composition sensitivity further.

\textbf{Scope conditions.} AGORA should be interpreted as a controlled sensitivity-analysis instrument rather than a behavioral replica of real public meetings. Results are conditional on persona specification and {LLM}-mediated critique generation, gate/objective settings, and benchmark-scale evaluation.

\section{Conclusion}
We presented AGORA, a framework that isolates how meeting composition propagates into transit-planning outcomes by fixing network, demand, and solver while systematically varying who participates and how deliberation is structured. Three findings emerge across two benchmark instances. First, aggregate composition sensitivity is bounded, with no statistically significant profile effects---yet directional analysis reveals that representative sampling tends to produce lower fairness disparity and tail risk (RQ1). Second, the no-revision control produces zero composition variation, identifying critique-mediated deliberation as the mechanism through which composition influences outcomes (RQ2). Third, on Mandl, governance gates compress cross-profile variance and reduce tail-risk differences; on Mumford0, low acceptance shows that gate thresholds must be calibrated by instance (RQ3). Together, these results suggest that \emph{process design mediates composition effects} more than composition itself drives outcome variation, and that representative synthetic populations combined with governance gates provide a robust default for {LLM}-mediated participatory planning.

Current transit engagement relies on one-directional comment collection from uncontrolled samples \cite{mbta2022bnrd}. AGORA's contribution is not to replace such processes but to provide a controlled testbed for studying how process design choices interact with composition effects, a question that observational data from real hearings cannot answer. The results reframe participation bias from an uncontrollable input problem to a tractable process-design problem: even when representative attendance cannot be guaranteed, well-structured deliberation and governance criteria can substantially reduce the sensitivity of planning outcomes to who is in the room.

Future work will focus on external validation: calibration against observed participation records, integration with real agency engagement workflows, and application to real-world instances where greater spatial and demand heterogeneity should amplify composition effects.

\section*{ACKNOWLEDGMENTS}
This work was supported by the East Japan Railway Company (JR East), the Kwanjeong Educational Foundation, and the Chishiki-AI SCIPE Graduate Fellowship.

\bibliographystyle{IEEEtran}
\bibliography{references}
\end{document}